\documentclass{edp-conf}
\usepackage{graphicx}
\begin{document}
\TitreGlobal{SF2A 2001}
\title{Galaxy and star formation up to z$\sim$ 1}

\runningtitle{Galaxy and star formation up to z$\sim$ 1}
%
\author{F. Hammer and H. Flores}\address{GEPI, Observatoire de Paris}

%
\maketitle
\begin{abstract} 
Star formation history shows a gradual decline since the last 8-9 Gyr
(z=1). The bulk of present-day stellar mass and metal content was
formed at redshifts lower than 2-3, which is consistent with a
hierarchical scenario of galaxy formation. Observations of galaxy
evolution during the last 2/3 of the Universe age could be done in
great details, and provide numerous insights on the origin of the
Hubble diagram. To some extent, the evolution and formation of
present-day elliptical and spiral galaxies could be related to the
most rapidly evolving galaxy populations at z$\sim$ 1, namely the
luminous IR galaxies and the luminous compact galaxies.
\end{abstract}
%
\section{Introduction}
Analysis of the age-metallicity relation in the solar neigborhood by
Twarog (1980) indicated higher star formation rates in the past by
factor 2-3, when the disk was from 1/3 to 1/2 of its present
age. Hubble Space Telescope studies of Local Group dwarves have
revealed a surprisingly large variety of evolutionary histories (see
e.g. Grebel, 2000).  Rapid technological developments in the last
decade have opened a new era for galaxy evolution studies: galaxies
can be detected at the earliest epochs of the Universe and their
evolution can be directly compared in different redshift
slices. Combination of nearby and distant Universe studies, is likely
the best way to understand how (and when) stars and metal content were
formed in the Universe.

The Canada France Redshift Survey (CFRS, Lilly et al, 1995) have
probed the evolution of luminous ($L^{\star}$) galaxies up to z=
1. The fraction of star forming galaxies rapidly increases with the
redshift: more than 50\% at z$>$0.5 have $W_{0}(OII)>$15\AA~ (Hammer
et al, 1997), which should be compared to 13\% locally (Zucca et al,
1998). This observed trend is followed by a variation of the average
rest-frame $(U - V)_{AB}$ color from 1.6 (Sab color) at z$\sim$0, to
1.3 (Sbc color) at z=0.5 and 0.7 (Sdm-Irr color) at z=1.
\section{Star formation history up to z=1}
The increase in star formation with redshift has been quantitatively
estimated by Lilly et al (1996) from the rest frame 2800\AA~
luminosity, whose comoving density evolves as rapidly as $(1 +
z)^{3.9\pm0.75}$. This value is provided after assuming a constant
slope of the galaxy luminosity function (GLF) in the $[0,1]$ redshift
range. This evolutionary change has been interpreted as due to a large
decrease of the star formation by a factor 10 from z=1 to z=0 (Madau
et al, 1996). This factor could be revised downward to $\sim$ 4, as
found by Cowie et al (1999), a large part of the discrepancy being
related to uncertainties on the local SFR density value.

However, optical studies of rest-frame UV light from distant galaxies
do not tell us all the story.  A large fraction of the UV luminosity
could be absorbed by dust and then reemitted in the far IR. The effect
is particularly significant for the luminous IR galaxies which
dominate the upper end of the bolometric GLF. Using CFRS data, Hammer
and Flores (1998) have shown that UV and [OII]3727 luminosities are
not correlating with $H\alpha$ luminosities, while they correlate well
together. This suggests that extinction is a major source of
uncertainty for determining the star formation history.

From a CFRS follow-up study with ISOCAM and VLA, Flores et al (1999)
have provided the first estimate of the SFR density at z$\le$ 1 which
was missed by optical measurements because of dust. They conclude that
4\% of the field galaxies at z$\le$ 1 are strong and heavily extincted
starbursts with SFR from 50 to 200 $M_{\odot}yr^{-1}$. This sparse
population is responsible for almost a third of the global star
formation density at z $\sim$ 1. Assuming a non evolved IR GLF, the
resulting global star formation density is 2$\pm$0.5 times higher than
derivations from UV measurements (Figure 1). The SFR density increases
by a factor ranging from 5 to 10 from the present epoch to z=1. The
uncertainties related to that study are coming from: (1) those related
to local values; (2) the ambiguity about the source of IR emission in
some luminous galaxies (Seyfert2); (3) a possible evolution of the GLF
lower end; and (4) a possible evolution of the IMF.
The observed SFR density decline since z=1 is mostly due to two
rapidly evolving populations (Figure 2), namely the luminous IR
galaxies (LIRGs, Flores et al. 1999) and the luminous compact galaxies
(LCGs, Guzman et al. 1997). LIRGs are large and relatively massive
starbursts often found in interacting systems. LCGs are an emerging
galaxy population at z$\ge$ 0.5, representing $\sim$ 30\% of the
galaxy population, while they have very few counterparts in the local
Universe.
\begin{figure}
\includegraphics[width=0.6\textwidth]{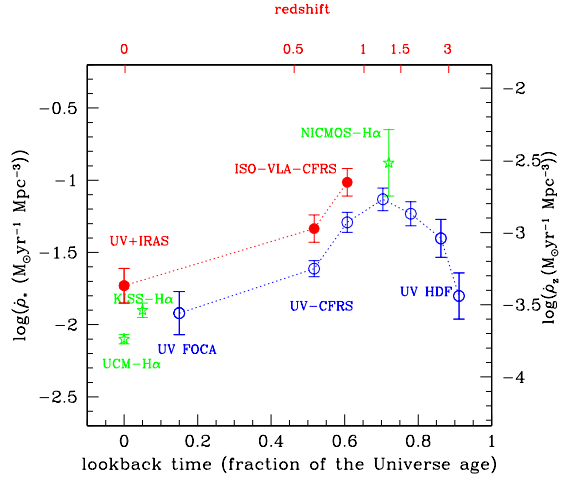}
  \caption{ {\small Observed metal production and star formation
  densities at different look-back times (from Hammer, 2000).  SFR
  estimates are assuming a Salpeter IMF from 0.1 $M_{\odot}$ to 100
  $M_{\odot}$. Flores et al points (filled circles, labeled
  ISO-VLA-CFRS) show a global SFR density (contribution of both UV and
  IR photons), two times higher than the values previously derived
  from UV (open circles).  Highest z estimates at rest-frame UV are
  from Connolly et al (1997) and from Madau et al (1996, HDF, open
  squares) and have not been corrected for extinction. Estimates from
  $H\alpha$ luminosities (stars) are from Gallego et al (1995; UCM),
  Gronwall (1998; KISS), and also from few galaxies at z$\sim$ 1.25
  (Yan et al, 1999; NICMOS).}}
\label{figure1}
\end{figure}  
\section{Testing scenarii of galaxy formation}
A simple integration of the global star formation leads to the
 formation of half of the present-day stars since z=1-1.5 (Flores et
 al, 1999; Madau et al, 2001; Franceschini et al, 2001). At z= 3,
 less than a third of the present-day metal content was formed, even
 accounting for large dust corrections (scenario of a constant star
 formation density beyond z$\sim$1).  This favors a scenario in which
 the bulk of the metal content in the Universe is formed at
 relatively recent epochs (z$\le$ 2), supporting a hierarchical
 scenario, i.e. that massive galaxies are gradually built up through
 the coalescence of smaller building blocks. The above considerations
 could challenge the assumption that all massive galaxies were formed
 through a primordial collapse at the earliest epochs (z$\ge$ 3),
 because most of the metal are locked into massive bulges (Fukugita
 et al, 1998).

When massive ellipticals (E/S0) have been formed ? Bernadi et al
(1998) found a tight color-$\sigma$ relation for a sample of
early-type galaxies in the field as well as in galaxy clusters. Such a
tightness demands a high degree of synchronicity in their star
formation histories, that is naturally accounted for by pushing back
to early times (z$\ge$ 2) most of their star formation (Bower et al,
1992; Renzini, 1999). A large fraction of massive ellipticals lie in
clusters or in groups, and having most of their star formation at high
redshifts could be accomodated in hierarchical models, because
clusters form out of the highest peaks in the primordial density
fluctuations (Kauffmann and Charlot, 1996). However, in low density
environments, the model predicts much recent epochs for their star
formation, including through mergers of dissipative gas rich disk
galaxies.  It has been suggested that ellipticals could be shared in
two classes, one class comprising low to moderate mass and luminosity
lenticulars and ellipticals, the other the most massive ellipticals
often found in groups or clusters (Faber et al, 1997). Both classes
occupy different part of the fundamental plane, and might have
different star formation history. Genzel et al (2001) have suggested
that ultra-luminous IR galaxies (ULIRG), which are found in merging
systems, are progenitors of the less massive ellipticals. At least
20\% of ellipticals could be formed during such events since z=1,
according to the number density of colliding giant disks detected by
ISO (e.g. Hammer, 2000).  
Another constraint to the epoch of field
E/S0 formation is provided by their number density 8-9 Gyr ago (z= 1):
it is comparable to that found locally (Schade et al, 1999), but the
result is limited by small number statistics.  Moreover, about 30\% of
morphologically selected ellipticals show signs of star formation
revealed by their colors and emission lines, and could not be
considered as non-evolved ellipticals as those seen today. From a
recent study of 145 E/S0 at z$\le$ 1, Im et al (2001) rule out that
more than 50\% of the ellipticals can be formed since z=
1. Present-day observations then suggest that from 20\% to 50\% of
ellipticals, mostly the less massive ones, were forming the bulk of
their stars at recent epochs (z$\le$ 1).
\begin{figure}
  \caption{ (ATTACHED JPEG FILE) {\small Example of the main contributors of the star
 formation history, i.e. a LCG ($z=0.77$) and a LIRG ($z=0.65$)
 detected by ISO. {\it Top} I band HST imagery.  {\it Bottom}
 VLT/FORS1 spectra (3 hour exposure, Hammer et al, 2001) which reveal
 the important absorption line system and strong emission lines. In
 the Right panel, the continuum and absorption lines are well
 reproduced by a combination of B, A, F, and G stars (dot-dashed line,
 Gruel et al, in preparation). }}
\label{figure2}
\end{figure}  
What is the star formation history of disk galaxies ? Modelling of the
Milky Way (see Boissier and Prantzos, 1999) as well as the Schmitt
law for disks (Buat, 1992; Kennicutt, 1998) argue in favor of a
rather long duration (3-7 Gyr) for the formation of the bulk of
their stars. The density of large disks ($r_{disk}\ge$
3.2$h_{50}^{-1}$kpc) is found to be the same at z=0.75 than locally
(Lilly et al, 1998). Having most the disks already in place at
z$\sim$ 1 is also supported by the apparent non-evolution of the
Tully Fischer relation as reported by Vogt et al (1998). Above
studies are however limited, either by model assumptions ( Boissier
and Prantzos model assumes no galaxy interaction), or by small
number statistics (Lilly et al and Vogt et al studies) or by a
somewhat unadapted observational set-up (slits not aligned to the
galaxy major axis in the Vogt et al study).

Clues for a high redshift formation of disks -and then a gradual disk
  formation during the last 8 Gyr- have been derived on the sole basis
  of optical observations. Conversely to that, IR observations reveal
  that a noticeable fraction of disks show recent and rapid
  evolutions, probably because of galaxy interactions.  Deep mid-IR
  counts show a strong evolution (Elbaz et al, 1999), mostly related
  to z$\le$ 1.2 disks, often found in interacting systems. About 30\%
  of the large disks at z=0.5-1 are IR luminous, and experience strong
  episods of star formation at rates of several tens of solar mass per
  year, exceeding by far estimates from rest-frame UV light. It is
  still unknown what is the fraction of z$\sim$ 1 disks which are
  disrupted during merging events. Galaxy interactions and extinction
  related effects could also severely affect dynamical measurements
  such those from the Vogt et al (1997) study.

 How spiral galaxies were forming ? What they look like during their
 early stages of formation ? Hammer et al (2001) have suggested that
 luminous compact galaxies (LCGs) were indeed progenitors of disks
 comparable or smaller to the Milky Way. LCG spectra (Figure 2)
 reveals strong metallic absorption lines combined with intense
 emission lines: their main stellar population are evolved stars with
 metal abundances comparable to those of Milky Way's bulge, although
 they are sites of intense star formations. In the Hammer et al 's
 interpretation, star formation was firstly occuring in bulges through
 merging of smaller entities often revealed by deep HST imagery, while
 faint surrounding extends of LCGs correspond to low surface
 brightness disks. LCGs are very enigmatic objects and understanding
 their nature needs more investigations: their sizes are similar to
 those of local dwarves, although they are up to one hundred times
 more luminous than dwarves.  Their kinematics from optical emission
 lines revealed low velocities, suggesting low masses comparable to
 those of present day dwarves (Koo et al, 1995; Guzman et al, 1997),
 although merging and large extinctions could severely affect the
 interpretation.
    
\section{Conclusion}
Strong events of star formation and important changes in the galaxy
populations were occuring during the last 8-9 Gyr. Current or
soon-coming instrumentation can provide details studies of z$\sim$1
galaxy properties, aiming at firmly establish the origin of the Hubble
diagram. Several progresses are required to understand how galaxies
were formed and to recover the universal star formation history,
including (this is not an exhaustive list):
\begin{itemize} 
\item an accurate estimate of the universal stellar
 mass density and its evolution from either galaxy dynamics
 measurements and/or a better calibration of stellar mass from near-IR
 luminosity measurements 
\item dynamical studies of a large number of
 merging systems in the past Universe, to investigate the nature of
 their by-products 
\item multi-wavelength analyses of star formation
 at all redshifts, which accurately account for dust-enshrouded star
 formation events 
\item proper measurements of gas dynamics in violent
 events of star formation, from measurements at far-IR rest-frame
 wavelengths, and/or a careful account of the selection effects
 related to extinction in optical measurements 
\item a better
 understanding of how stars are forming in a primordial medium, as
 well as improvements in modelling low metallicity stellar populations
 \item accurate stellar mass functions in various environments,
 including in low abundance medium, in powerful starbursts and in
 massive bulges 
\item a better knowledge of the Type II AGN energy
 output, because they significantly contribute to the upper end of the
 bolometric GLF.  
\end{itemize}
 Several of the above challenges require significant progresses
 related to the stellar physics. VLT has a major role to play,
 especially with its soon-coming instrumentation. Velocity fields of
 galaxies -including those in interaction- will be provided by GIRAFFE
 with its 3D spectroscopic mode and at intermediate resolution (R$\ge$
 5000, $\delta$v $\sim$ 30 km/s).  At the focus of an 8 meter
 telescope and from its large multiplex, GIRAFFE will be unique for
 investigations of the low abundance stellar properties by observing a
 large number of stars up to the Magellanic Cloud and further
 away. Star formation histories of Local Group galaxies will be
 carefully analysed by measuring the metal abundance of their giant
 stars. Deepest spectroscopic surveys by VIMOS will provide invaluable
 insights of the GLF well below $L^{\star}$ up to z=1.
  After its launching, SIRTF will be unique for determining the IR
  luminosity function up to large redshifts (z=2-3), and then will
  firmly establish the fraction of star formation occuring in very
  dusty environments. Further important progresses are also expected
  during the next decade. At sub-mm wavelengths, ALMA will be able to
  detect the earliest events of dust-enshrouded star formation,
  because, at high redshift, the 100 $\mu$m thermal peak progressively
  enters the sub-mm window, which almost compensates the cosmological
  dimming. It will also provide accurate measurements of the gas
  dynamics at sub-arcsecond scales, independently of extinction
  effects. At near-IR, the NGST will fully investigate the optical
  emission properties of highly redshifted galaxies. Further
  progresses are however needed for improving the spatial resolution,
  because 1kpc at z$\ge$ 0.7 represents 0".1, twice the resolution the
  NGST (diffraction limit of a 6.5m at 2.2$\mu$m). Development of
  adaptive optics from the ground, including from multi-conjugate
  analyses, is a very promising way and could generate instruments for
  the 2nd generation at VLT. For example, FALCON could complement the
  NGST by studying velocity fields in distant starbursts.  Such
  developments are a prerequisite for the next generation of extremely
  large telescopes (ELTs with diameters $\ge$ 20 meters). From their
  large collecting areas and their image quality (assumed to be
  restored near their diffraction limits), ELTs such as NG-CFHT, CELT
  or OWL, would allow a new giant step in our understanding of the
  distant Universe.


\end{document}